# Hyperbolic Dispersion Dominant Regime Identified through Spontaneous Emission Variations near Metamaterial Interfaces

Kwang Jin Lee [a,b]*, Yeon Ui Lee [a,b], Sang Jun Kim [c], Pascal André [b,d]*

[a] Department of Physics, Quantum Metamaterials Research Center, Ewha Womans University, Seoul 03760, South Korea ; [b] Department of Physics, CNRS-Ewha International Research Center, Ewha Womans University, Seoul 03760, South Korea ; [c] Ellipso Technology Co. Ltd., 358 Kwon Gwang-ro, Paldal-gu, Suwon, South Korea ; [d] Elements Chemistry Laboratory, RIKEN, Hirosawa 2-1, Wako 351-0198, Japan

**Abstract**
Surface plasmon polariton, hyberbolic dispersion of energy and momentum, and emission interference provide opportunities to control photoluminescence properties. However, the interplays between these regimes need to be understood to take advantage of them in optoelectronic applications. Here, we investigate broadband variations induced by hyperbolic metamaterial (HMM) multilayer nanostructures on the spontaneous emission of selected organic chromophores. Experimental and calculated spontaneous emission lifetimes are shown to vary non-monotonously near HMM interfaces. With the SPP and interference dominant regimes.
With the HMM number of pairs used as the analysis parameter, the lifetime is shown to be independent of the number of pairs in the surface plasmon polaritons, and emission interference dominant regimes, while it decreases in the Hyperbolic Dispersion dominant regime. We also show that the spontaneous emission lifetime is similarly affected by transverse positive and transverse negative HMMs. This work has broad implications on the rational design of functional photonic surfaces to control the luminescence of semiconductor chromophores.

## 1. Introduction

Surface and interface design nurtures very active research fields with various applications such as optical sensing,[1-5] optoelectronic devices,[6-8] chirality,[9,10] magnetism,[11-13] and surface-enhanced Raman spectroscopy.[14-16] Controlling the structure of an environment is known to impact on photophysical properties of semiconductors. The effects can be revealed through alterations of energy and charge transfers,[17-19] spectral shifts and amplitude variation of both absorption and transmission,[20-23] as well as polarization,[24] and emission amplitude.[25-28] In this context, engineering spontaneous emission rate with elaborated nanostructures is of particular interest and the effect of various structures has been explored including confinement,[28-30] plain metal films and metal gratings,[31-35] cavities,[36-39] antennas,[40,41] photonic crystals,[42] nanoholes,[27] metal and hybrid nano-particles,[43-45] to name but a few.

Focusing on metal structures, important and noticeable examples of surface alterations of photophysical properties find their origins in Drexhage's early experiments. Interactions between a transient dipole and a metallic mirror have been discussed in terms of interference between the radiated field and the reflected radiation from the mirror.[46-48] This leads to a non-monotonous variation of the normalized decay rate as a function of the distance from the mirror interface.[33,49] Interference between the directly emitted field and that reflected at the metal surface results in an oscillating behavior. Such an approach is relevant to resonant energy transfer,[18] or scattering near metal interfaces,[50-52] and it was recently revisited with acoustic waves.[53] Alternatively, photonic density of state (PDOS) can be used as complementary formalism to describe luminescence decay rates near metal surfaces.[33] Based on Fermi's golden rule, both plasmonic coupling and interference contribute to PDOS. The effect of the former can be assessed by calculating the Purcell factor, which is strongly affected by plasmonic modes, when quantum emitters interact with plasmonic structures. In the near field, short distance range up to 20 ~ 30 nm, the emission characteristics are significantly altered due to the plasmonic coupling with the dipole radiations. In contrast, the effect of the latter takes over in the far-field through emission interference.

Beyond single metal mirrors, more complex metal structures have been explored to control the photophysical properties of semiconductors. Among them, nanoplasmonic metamaterials,[54,55] photonic hypercrystals,[56] and metamaterials [57-72] have drawn a lot of attention in recent years because of their unusual electromagnetic properties. Within this latter class of structures, hyperbolic metamaterials (HMMs) are extremely anisotropic subwavelength nanostructures, which can be made of rods immersed into a dielectric environment or of stacked multilayers.[73] In contrast with conventional dielectric media showing elliptic dispersion, HMM structures present a hyperbolic dispersion (HD), which originates from the opposite signs of the dielectric permittivity tensors of its two main components.[57] This results for instance in supporting high-$k$ states since it was demonstrated that bulk propagating waves with large wave vectors in periodic multilayer HMMs originates from coupling of SPPs in the individual metal layers.[58] Therefore, HMMs have been intensively studied in the context of negative refraction,[59,60] far-field sub-wavelength imaging,[61-63,74] plasmon polaritons,[75] and spontaneous emission rate alterations.[64-72,76] However, while the HMM's large PDOS was shown to increase the radiative decay rate of fluorescent emitters, it remains unclear how the crossover between plamonic, HD and interference effects occurs and what parameter would be best used to formalize possible crossovers from one dominant regime to another.

## 2. Results and Discussion
### 2.1. Optical Spectroscopy

In this work, we investigate these issues with a set of organic chromophores, namely 4,4'''-bis[(2-butyl octyl)oxy]-1,1':4',1":4",1'''- quaterphenyl (BBQ), Coumarin 460 (C460) andN,N-Di(1-octylheptyl)-perylene-3,4,9,10-tetra carboxylic diimide (PerDi), which chemical structures are presented in Figure 1A. They were selected to be soluble in dichloromethane (DCM), to be miscible with polymethylmethacrylate (PMMA) and to span across the UV-vis spectral range. Dispersed at a low concentration in a PMAA solution, they were spincoated on top of four-pair HMM structures. Two polymer thicknesses of 15 and 35 nm were used. These are comparable to the thicknesses used in previous HMM studies,[65-67, 77] and within the experimental precision both polymer film thickness displayed the same PL decays. The structure of the substrates, which this work focuses on, are made of 10 nm thick Ag and $Al_2O_3$ alternative thin films deposited on fused silica (FS). Focusing on the nanophotonic control of the dye emission, we tuned the thickness of the top $Al_2O_3$ cover, which is defined as distance $d$, and is schematically illustrated in Figure 1B. Seventeen different samples were prepared with the top $Al_2O_3$ cover thickness ranging from 10 nm to 1 μm. The effective dielectric constants $\varepsilon_{x,y}$ and $\varepsilon_z$ of such HMM structures can be calculated based on the effective medium approximation and display a transverse positive-to-negative transition near 375 nm as shown in Figure 1C. The optical properties of the resulting samples were then characterized experimentally by a combination of steady state and time-resolved fluorescence spectroscopies. The results were analyzed with a theoretical approach relying on the combination of the invariant imbedding method,[78, 79] and the finite-difference time-domain (FDTD) method.[80, 81] With these tools, we calculated the spontaneous emission rate and the Purcell factor as a function of the distance between the dye-polymer film and the top metal layer of the HMM substrates. We then identify SPPs, HD and interference emission interference dominant regions as a function of the emitters HMM substrate separation. We also show that a dimensionless parameter independent of the emission spectral range is useful to identify these three regions.

Figure 2ABC1 shows the steady state normalized absorbance and fluorescence spectra of the three organic chromophores. The absorption and emission spectra agree well with the literature. The maximum absorbance wavelengths are 309 nm, 371 nm, and 524 nm for BBQ, C460 and PerDi, respectively. Solubilized in DCM, their photoluminescence (PL) spectra cover 350-470 nm,

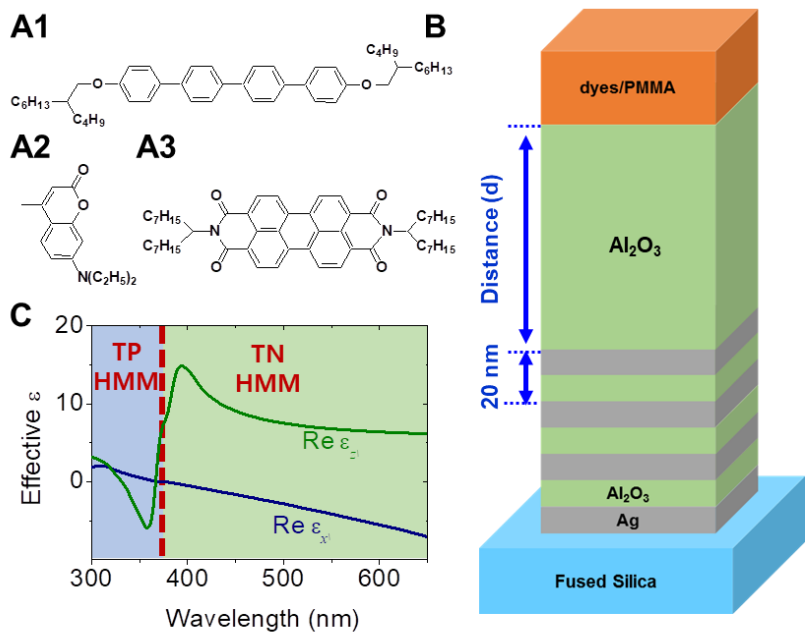

**Figure 1.** (*A*) Chemical structure of BBQ (*A1*), C460 (*A2*) and PerDi (*A3*). (*B*) Schematic of the 4-pair HMM-PMMA:dye samples. (*C*) Effective permittivity of the HMM used in this study. The dashed line is 375 nm boundary wavelength between transverse positive (TP) HMM and transverse negative (TN) HMM.

405-485 nm and 520-630 nm, respectively. Immersed in a PMMA matrix, the steady state normalized fluorescence spectra of the three chromophores are slightly broader which is due to the vibronic interactions and inhomogeneous dielectric environment provided by the PMMA solid solution. A weaker effect is observed with PerDi than with BBQ and C460. The photoluminescence spectra of the three chromophores are unaffected, when spin-coated either on FS, or on HMM substrates (see Figure S1).

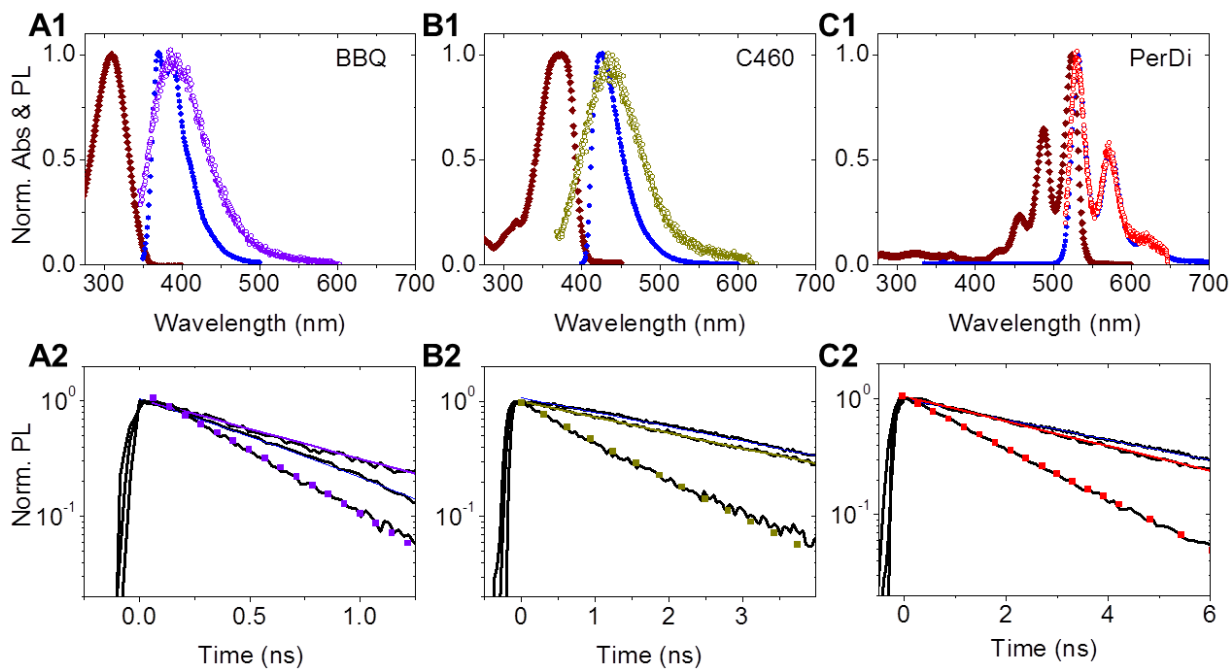

**Figure 2.** Photophysical properties of a selection of chromophores: (*A*) BBQ, (*B*) C460 and (*C*) PerDi. (1) Steady state normalized absorbance (OD, ♦ ) and emission (PL, ●) of the dyes in DCM solutions (filled symbols) and their emission in PMMA matrix from a thin film spin-coated on top of fused silica substrates (empty symbols). (2) Time resolved photoluminescence decays in DCM (thin solid line fit), in PMMA on fused silica (thick solid line fit) and in PMMA on HMM substrates (filled symbol fit, ■). The excitation wavelength of BBQ and C460 is 325 nm, while PerDi's is 470 nm.

Streak camera measurements were performed to monitor the evolution of the spontaneous emission lifetimes for BBQ, C460 and PerDi. The resulting streak images of three chromophores in PMMA on FS and HMM are shown in Figure S2 to S4. Integrations of the streak camera signal along the time axis within the emission spectral regions are shown in Figure 2ABC2. All the time resolved PL spectra, regardless of the solvent, matrix and substrate conditions, display a mono-exponential decay, indicating that there is no significant aggregation in either solution or film. The PL lifetime values are summarized in Table S1. Within 10 %, the emission dynamics of three chromophores are identical when measured in DCM and in PMMA on FS. As expected from the literature, dyes in PMMA spincoated on HMM structures present a noticeably faster PL decay. Decay time of each sample in the presence of HMM significantly decreases over the whole spectral range herein considered. This acceleration of the spontaneous emission on HMMs is then confirmed to be a broadband effect.

## 2.2. Numerical Simulations

Theoretical calculations of the normalized spontaneous emission lifetimes at three wavelengths relevant to BBQ, C460 and PerDi ($\lambda_{PL}$ = 365 nm, 450 nm and 535 nm) are shown in Figure 3A. The calculations were performed using the following equations [33]:

$$\tau_\parallel(\zeta) = \tau_0 \left[ 1 + \frac{3}{4} q \, \text{Im} \left( \int_0^\infty [(1-u^2) r_p + r_s] e^{-2w\zeta} \frac{u}{w} du \right) \right]^{-1} \quad (1)$$

$$\tau_\perp(\zeta) = \tau_0 \left[ 1 - \frac{3}{2} q \, \text{Im} \left( \int_0^\infty r_p e^{-2w\zeta} \frac{u^3}{w} du \right) \right]^{-1} \quad (2)$$

where $\tau_\parallel$ and $\tau_\perp$ are the decay times of radiative dipoles having parallel and perpendicular orientations, respectively, when compared with the multilayered HMM structures. For sake of simplicity, we define $\zeta = d \cdot k_d$ as the dimensionless parameter characterizing the minimum distance of the transient dipoles from the last HMM Ag layer (i.e. spacer thickness) in terms of wavevectors. $\tau_0$ is the intrinsic PL lifetime of the emitter. $q$ is the radiative quantum efficiency of the emitter and is set to a constant value independent of the distance, $u = (k_d)_x/k_d$ is the wavevector ratio, and with $w = -i(1-u^2)^{1/2}$. $k_d$ and $(k_d)_x$ are the wavevectors inside the $Al_2O_3$ spacer layer, its amplitude and its component along the propagation and tangential axis, respectively. They are given by $2\pi n_d/\lambda_{PL}$ where $n_d$ is the refractive index of spacer layer and $\lambda_{PL}$ the emission wavelength of the radiative dipole. Finally, $r_s$ and $r_p$ are the reflection coefficients for s- and p- polarized electromagnetic waves, respectively. As detailed in the SI, in order to obtain the exact values of $r_s$ and $r_p$ of the HMM structures, we employed the invariant imbedding method for the electromagnetic wave propagation in the HMM structures. The equations of the reflection coefficients $r_s$ and $r_p$ are obtained and integrated for each thickness of the HMM top dielectric layer.[72, 78, 79, 82] With this approach, any dipole orientation, defined as $\gamma$, is considered as a linear combination of parallel and perpendicular dipole components. The decay time for any orientation is

$$\tau_\gamma = \gamma \tau_\parallel + (1-\gamma) \tau_\perp \quad (3)$$

Figure 3A shows the normalized spontaneous emission decay times $\tau/\tau_0$, i.e. the ratio of the spontaneous emission decay time of chromophores located on top of HMM and on top of FS substrates, inverse Purcell factor, as a function of both $\zeta$ (vertical axis) and $\gamma$ (horizontal axis), where blue and red represent slower and faster relative radiative lifetime, respectively. The three wavelengths, $\lambda$ = 365, 450 and 535 nm, correspond to the emission maximum of BBQ, C460 and PerDi, respectively. The overall behavior of $\tau/\tau_0$ is the same for the three chromophores under consideration: near the substrate, small $\zeta$ value, a shorter emission lifetime occurs, it then rises to a maximum value larger than one, and eventually displays periodic and damped oscillations with $\zeta$. $\tau/\tau_0 =1$ is eventually reached far from the HMM structure. With an orientation of the dipole perpendicular to the substrate, the damping takes place very close to the HMM structure. However, when the radiative dipoles become parallel to the substrate the amplitudes of the $\tau/\tau_0$ oscillation are larger and remain noticeable far from the HMM structure. We note that, with the emission wavelength of the transient dipole, the maximum value of $\tau/\tau_0$ increases and the green zone near the HMM structure appears to stretch over a wider $\zeta$ range.

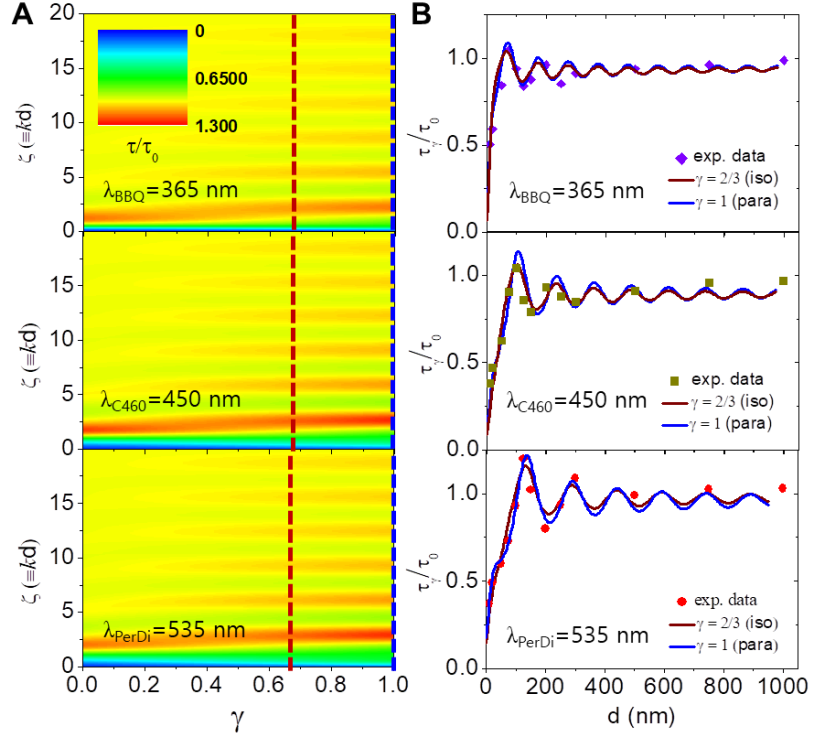

**Figure 3.** Effect of HMM spacer on spontaneous emission lifetime. (A) False-color plot of the calculated spontaneous emission lifetime as a function of the dimensionless distance ($\zeta = 2\pi d \cdot n_d/\lambda_{PL}$) and the dipole orientation ($\gamma$) for three wavelengths (365 nm, 450 nm, 535 nm) characteristic of the emission of the BBQ, C460 and PerDi emitters, respectively. The vertical dashed lines materializes isotropic (red) and parallel (blue) orientations of the emitting dipoles (B) Normalized spontaneous emission lifetimes for three emitters as a function of distance : experimental data (symbol), theoretical calculations (lines) where brown and blue lines correspond to $\gamma=2/3$ and $\gamma=1$, respectively.

Figure 3B presents cross-sections of Figure 3A for dipoles with isotropic and parallel orientation ($\gamma$ = 2/3 and 1) as a function of dimensionless distance, $\zeta$. Figure 3B shows a good agreement of the calculated and experimental data obtained with BBQ, C460 and PerDi. The positions of the first $\tau/\tau_0$ maximum of three chromophores are 75, 100 and 125 nm, respectively. The $\tau/\tau_0$ periodicities are 105 nm, 133 nm and 155 nm, which is directly related with the emission wavelength of the radiative dipole and corresponds to $\lambda/(2n_d)$, as illustrated in SI. Two differences between the three chromophores are nonetheless noticeable. First, BBQ and PerDi experimental results are more consistent with the calculations completed for $\gamma$ = 1 rather than $\gamma$ = 2/3. In contrast, the C460 experimental data are more consistent with the $\gamma$ =2/3 calculations. As further illustrated in Figure S5, this suggests that the chromophore orientations differ slightly in the PMMA matrix.

The smaller sizes and more compact shapes of the molecules lead to more isotropic orientations. In addition, we note that for short distances from the HMM structure, $\tau/\tau_0$ presents a kink before reaching its maximum values. As displayed in Figure S6, positive and negative kinks are observed for BBQ and PerDi, respectively. These two chromophores also present the shortest and longest emission wavelength. To focus on this aspect, PerDi was selected because this chromophore is associated with the largest amplitude of $\tau/\tau_0$ value variation and that its peak value is located the furthest away from the HMM structure. This situation allows an easier modulation of the dielectric spacer thickness.

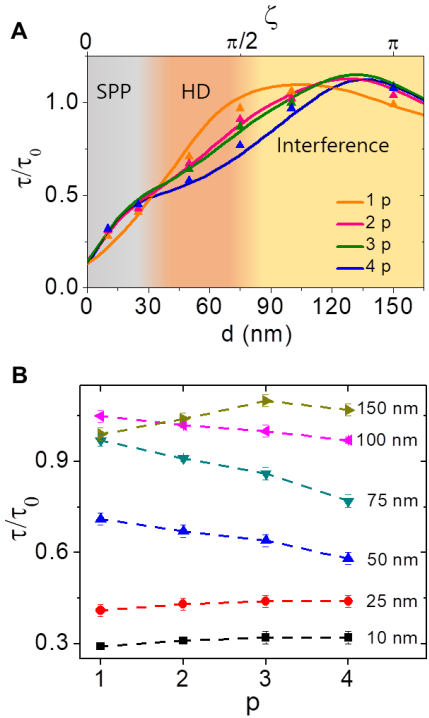

**Figure 4.** Emission lifetime ratio of transient dipoles at 535 nm. (A) Calculations (solid line) and experimental data of PerDi (symbols) of normalized fluorescence lifetime as a function of distance from HMM structure composed of 1 to 4 Ag/Al$_2$O$_3$ pairs and as a function of the dimensionless parameter $\zeta$. The colored background illustrates the crossover from one dominant regime to another. (B) experimental data as a function of the number of pairs of the HMM structure.

Figure 4A displays experimental data and calculations of $\tau/\tau_0$ as a function of distance from substrates composed of a different number of Ag/Al$_2$O$_3$ pairs with dipoles emitting at $\lambda = 535$ nm. The curves correspond to cross-sections of the false-color plots shown in Figure S7 of the emission lifetime ratio calculated as a function of distance and orientation parameter for 1~4 Ag/Al$_2$O$_3$ pairs. For each dielectric spacer thickness, the experimental data points are in qualitative good agreement with the calculated $\tau/\tau_0$ curves for each pair. For any number of pairs the variation of $\tau/\tau_0$ with $d$ is non-monotonous. Overall, $\tau/\tau_0$ increases up to a maximum larger than one and decreases. This is consistent with the behavior of 4 $p$ HMM structures reported in Figure 2. Interestingly, the distance, at which this maximum occurs, is shorter for 1 $p$ and barely varies for $p \geq 2$. The first maximum distance of $\tau/\tau_0$ and maximum value of $\tau/\tau_0$ with $p$ are plotted in Figure S8A and B, respectively. The inversion of $\tau/\tau_0$ with the number of pair is associated with the transient dipole being closer or further away than the position at which the maximum value of $\tau/\tau_0$ occurs. Furthermore, the kink previously mentioned is not visible at $p = 1$, but it appears and strengthens with $p$. These observations suggest that beyond the generality of the present experimental and numerical observation several phenomena take place near the HMM interface.

For very small distances, transient dipoles excite propagating surface plasmon polaritons (SPP) in the metal layers located nearby. A partial transfer of energy into heat also occurs due to losses in metal. Through these processes, PDOS increases and the non-radiative decay rate increases.[82,83] As a result, the photoluminescence lifetime is strongly reduced.[66]

This is consistent with the experimental $\tau/\tau_0$ data obtained on top of 10 nm and 25 nm Al$_2$O$_3$ spacers, which present very small values well below unity. With a 10 nm Al$_2$O$_3$ spacer, the marginally lower $\tau/\tau_0$ value for 1 pair substrate shown in Figure 4A and B could be associated with the very high Purcell factor around 550 nm which as presented in Figure S9 which occurs coincidentally in the same spectral range as the emission of the PerDi. Similarly, the 25 nm Al$_2$O$_3$ spacer generates weaker nonradiative interactions, induced by plasmonic coupling and heat losses, than 10 nm Al$_2$O$_3$ spacer. It is then consistent that 25 nm spacer presents slightly larger $\tau/\tau_0$ values, which nonetheless remain relatively independent of the number of pairs. Importantly, we note that in the so-called SPP dominant regime, the reduction of PL lifetime is mostly affected by the top metallic layer, which means that in this dominant regime the multilayer structure does not have a strong impact on $\tau/\tau_0$. This is confirmed both experimentally and numerically in Figure 4. In Figure 4A, we note that the theoretical calculations of $\tau/\tau_0$ for 2 to 4 pairs present a kink and then a separation of the curves around a spacer thickness of ~30 nm. As the SPP dominant regime corresponds to $\tau/\tau_0$ being relatively independent of the number of pairs, this kink was then used phenomenologically to set the crossover, $d_{\text{SPP-HD}}$, between SPP and HD dominant regimes. This is illustrated in Figure 4A by the background color progressive evolution from grey to orange. In the present system and with a 535 nm illumination, the distance over which SPP effects dominate is typically of the order of $d \leq 30$ nm. We stress that the exponential decay of the SPP field is influenced by optical constants as well as the metal and dielectric thicknesses, so that the crossover is a function of the spectral emission.

The situation is quite different in the HD dominant regime where the multilayer structures of the substrate does have a relatively strong impact on $\tau/\tau_0$. As the distance increases, the relative impact of the SPP effect, essentially driven by the top sliver layer, decreases and the PDOS enhancement by multi-metal layer structures is more strongly altered, which leads to an overall increase of the radiative relaxation channels. This is in agreement with the separation of the 1~4p $\tau/\tau_0$ curves visible in Figure 4A and the larger $\tau/\tau_0$ values obtained with d = 50 nm and 75 nm and presented in table 1. As shown in Figure 4A and B, the experimental PL decay characteristic times shorten with the number of Ag/Al$_2$O$_3$ pairs. Again the experimental data are in qualitative agreement with the numerical calculations. This behavior is also consistent with the literature in which radiative channels are known to increase when the emitter is located nearby HMMs.[66] Furthermore, we note that in this HD dominant regime non-radiative energy transfer from the excited molecules to the metal could also occur through dipole-dipole interaction.[33]

It is known that at large distances from a single metal interface, an oscillating variation of $\tau/\tau_0$ with a $\lambda_{\text{PL}}/2n_d$ periodicity is imposed by an interference effect resulting from the superposition of the emitted light field and its reflection at the metal interface.[33]
The single metal film literature defines the distance at which an oscillating behavior become relevant as $d_{\text{Int.}} \geq \lambda_{\text{PL}}/4n_d$, which correspond to $\zeta_{\text{Int.}} = \pi/2$.[33] The dimensionless parameter $\zeta$ is for such substrates a unique formalisation parameter.

The crossover to the interference dominant regime clearly depends on both the emission wavelength, $\lambda_{\text{PL}}$, and the refractive index of dielectric spacer, $n_d$. In the present system using $\lambda_{\text{PL}} = 535$ nm and $n_d = 1.78$ leads to an interference dominant regime crossover around 75 nm. This distance was illustrated in Figure 4A by the color background progressive change from orange to yellow. By analogy, we also used $\lambda_{PL}/4n_d$ to identify a crossover to the interference dominant regime for p > 1. This is further justified by FDTD calculations showing that the periodicity of the electric field oscillation is relatively unaffected by the number of pair, as presented in Figure S10A. Table 1 and Figure 4B show that at $d$ = 100 nm the impact of the number of pair of the PL lifetime is relatively weak, while at $d$ = 150 nm, $\tau/\tau_0$ appears to increase slightly with the number of pairs. At very large distance from the HMM top metal interface, the amplitude of the $\tau/\tau_0$ oscillations with distance decreases while the effect of the number of pairs becomes negligible, again as shown in Figure S10B.

Illustrations of the distinct effects on SPP, HD and interference dominant regimes are given in Figure 4B and Figure S11. Whilst it is nearly impossible to isolate one effect from the others, experimental opportunities exist to separate partially the contributions associated with SPP, HD, and interferences effects. For instance, there is no HD contribution for 1p structures which are only affected by SPP and interference effects. Next, the spectral covered by both chromophores and HMM building materials can be used to quench SPP contributions. Figure S12A presents the dielectric constants of the Ag and whilst the real component of the Ag dielectric constant is $\varepsilon_R < 0$ at 365 nm, there is a weaker SPP contribution at this wavelength than at 450 nm and at 535 nm for which $\varepsilon_R$ values are 2.7 times and 4.6 times smaller, respectively, when compared with $\varepsilon_R$(365 nm). Figure S13 presents the calculated $\tau/\tau_0$ spacer curves for 1~4p for three emission wavelengths. At 365 nm, which has the weakest SPP contribution, it is noticeable that there is no $\tau/\tau_0$ independent regime close the multi-layered structure, i.e. no SPP dominant regime. Then, $\tau/\tau_0$(365nm) starts directly by being p-dependent and the 1st peak of $\tau/\tau_0$ is divided by ~2 when comparing 1p and 4p, i.e. the PL decay is twice as fast in the latter than in the former substrate structure. This is consistent with the expected HD dominant regime. In contrast, $\tau/\tau_0$(450nm) and display a relatively p-independent behaviour up to about ~25 nm and ~30 nm spacer thickness, respectively. While both kinks and $\tau/\tau_0$ curve dispersition with p are more noticeable at 535 nm than at 450 nm. These features confirm that a few options could be explored to discriminate SPP, HD and interferences contributions in multilayer structures. Follow-up investigations could then use metals with a negative real dielectric constant occurring at a larger wavelength, or probing at wavelength that do not support surface plasmon resonance in the metal layer, or select absorbing dielectric spacers to reduce interference effects, hence to be able to gain futher insight on $d_{SPP-HD}$, $d_{SPP-Int.}$, and $d_{HD-Int.}$.

**Table 1.** Experimental values of $\tau/\tau_0$ ratio for PerDi at selected distances ($d$) from the substrates composed of 1 to 4 Ag/Al$_2$O$_3$ pairs (# p).

| # p / $d$ | 1 p | 2 p | 3 p | 4 p |
|---|---|---|---|---|
| 10 nm | 0.29 ± 0.01 | 0.31 ± 0.01 | 0.32 ± 0.02 | 0.32 ± 0.02 |
| 25 nm | 0.41 ± 0.02 | 0.43 ± 0.02 | 0.44 ± 0.02 | 0.44 ± 0.02 |
| 50 nm | 0.71 ± 0.02 | 0.67 ± 0.02 | 0.64 ± 0.02 | 0.58 ± 0.02 |
| 75 nm | 0.97 ± 0.02 | 0.91 ± 0.01 | 0.86 ± 0.02 | 0.77 ± 0.02 |
| 100 nm | 1.05 ± 0.02 | 1.02 ± 0.01 | 1.00 ± 0.02 | 0.97 ± 0.01 |
| 150 nm | 0.99 ± 0.02 | 1.04 ± 0.02 | 1.10 ± 0.02 | 1.07 ± 0.02 |

To summarize our findings, in the present studies three distinct dyes covering a wide spectral range were used to identify the general behavior of transient dipoles near HMM type of structures and to evidence crossovers regimes dominated by SPP, HD and interference effects. The chromophore with the lowest transition energy and HMM substrates with different number of pairs were then used to gain futher insights. It was then shown that the SPP dominant regime is characterised by a *p*-independent $\tau/\tau_0$ behavior, the HD dominant regime is characterised by an acceleration of the emission decay with *p*, and the interference dominant regime eventually displays again p-independent $\tau/\tau_0$

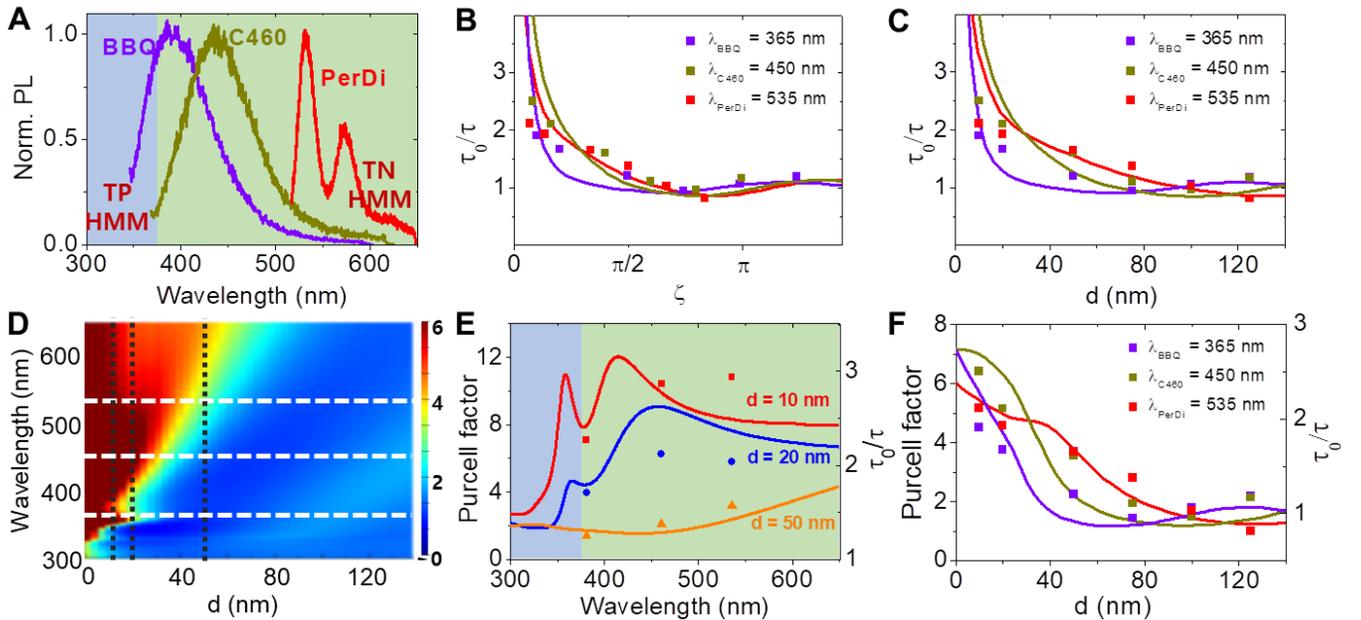

**Figure 5.** (A) Fluorescence spectra of three chromophores dispersed in PMMA matrices ; the vertical dashed line is the boundary wavelength between transverse positive (TP) and transverse negative (TN) of 10nm:10nm Ag:Al$_2$O$_3$ HMM structures. (B, C) Analytical calculations and experimental data of normalized spontaneous emission rates for $\lambda_{PL}$ = 365, 450 and 535 nm for three chromophores as a function of (B) the dimensionless parameter and (C) distance. (D) Pseudocolor plot of Purcell factor as a function of wavelength and distance from the HMM top metal layer. (E) Cross-sections of the Purcell factor as a function of wavelength taken at d = 10, 20 and 50 nm. (black dash line in D) (F) Purcell factor (solid line) as a function of distance for three wavelengths (365, 450, 535 nm, white dash line) and the inversed of normalized experimental decay rate $\tau_0/\tau$ (symbol) for the chromophores as a function of distance. (B) to (F) are for 4 Ag/Al$_2$O$_3$ pair HMMs.

periodic oscillations.

Compared to single metal film based system, the complexity of HMM types of structures is that they see not two but three overlapping contributions, and it is noticeable that one effect dominant regime does not exclude any impact of the two other effects. Whilst beyond the scope of the present study, we note that, in contrast of single metal film systems, the dimensionless parameter $\zeta$ can only be used as a guide, but not as a unique formalisation parameter to explore HMM structures impact on PL dynamics. $\zeta$ does not contains any dependence in $p$, which is not the case of HMM structure photophysical properties. Crossovers, $d_{SPP-HD}$, $d_{SPP-Int.}$, are emission wavelength dependent and a function of the HMM dielectric constants but their exact values, and the width over which they extend remain to be explored.

The final part of this work addresses the effect of the boundary between transverse positive (TP, where $\varepsilon_{x,y} > 0$ and $\varepsilon_z < 0$) and transverse negative (TN, where $\varepsilon_{x,y} < 0$ and $\varepsilon_z > 0$) of the Ag:Al$_2$O$_3$ HMM structures. For this reason, we now focus on the SPP, HD dominant regimes, by considering three chromophores and comparing results from optical and FDTD numerical calculations.

As illustrated in Figure 5A, the emission spectrum of BBQ lies across the TP and TN HMM spectral ranges, while those of C460 and PerDi lie in the TN HMM region only. Calculation and experimental results of inverse of $\tau/\tau_0$ as a function of the dimensionless parameter and the distance are shown in Figure 5B and 5C, respectively. Experimental data and theoretical calculations based only on the optical properties of the substrate show similar trends in both representations. Let's note that the color background in Figure 4A cannot be reproduced in Figure 5 as the crossovers are wavelength dependent. Across both the SPP and HD dominant regimes, a qualitative agreement is observed with longer (smaller) decay time (rate) for BBQ at 365 nm than for C460 at 450 nm and PerDi at 535 nm. This was confirmed with a fourth chromophore (2,5-diphenyloxazole) whose emission spectral range is slightly lower than BBQ's as shown in Figure S13. A qualitative agreement is observed between the experimental data and the optical reflection numerical approach, with for instance of the relative PL lifetime being systematically slower for emissions near the TP to TN boundary wavelength. This would be attributed to the different PDOS in the TP and TN HMM regions.

In order to assess this hypothesis, the Purcell factor (PF) was calculated by the FDTD method as a function of the distance from Ag/Al$_2$O$_3$ 4-pair HMMs and over the spectral range considered in this study. Figure 5D shows that the Purcell factor strongly depends on the distance and wavelength; maxima are located near the HMM interface while at any incident wavelength PF decreases with distance. For an easier reading, Figure 5E displays vertical cross-sections of Figure 5D (see the black dotted lines). In the Purcell factors plots, two peaks are clearly visible. The first around 360 nm corresponds to silver plasmon resonance. The second at longer wavelength corresponds to coupling between the metal layers. Overall, the Purcell factor is lower in TP than in TN HMM region, and it decreases when the dielectric spacer increases. The peak and shoulder shown for the $d$ = 10 nm and 20 nm curves, respectively, are associated with the surface plasmon resonance of Ag film. Since the Purcell factor is proportional to local density of state,[84, 85] Figure 5E is consistent with a previous numerical study showing that local density of state in TP HMM region is lower than that in TN HMM region.[85]

While the FDTD calculated Purcell factors are plotted on the lefthand side of Figure 5E the experimental $\tau/\tau_0$ data are plotted the right handside for comparison purposes. A qualitative agreement can be seen. The difference between the two sets of data is explained by the assumption of FDTD calculation that the emitter is a pure transient dipole whereas practically, organic fluorophores have intrinsic molecular nonradiative pathways.

We also plotted the cross-sections of the Purcell factor taken at $\lambda_{PL}$ = 365, 450 and 535 nm (horizontal white dashed lines) from Figure 5D and experimental data of normalized decay rate for three chromophores. The results shown in Figure 5F demonstrate that the Purcell factor and emission decay rate enhancement of organic chromophores in TP HMM region are smaller than those in TN HMM region.

In this work, we have provided two theoretical descriptions based on two different methods, analytical formalism combined with invariant imbedding method and Purcell factor using FDTD simulations, to compare with our experimental results. The trends observed with these three different approaches are consistent with one another, while the differences between two theoretical results could come from parameters such as the reflection coefficients obtained by the invariant imbedding method, mesh step size in FDTD simulations (see Method in SI). In Figure S15, we calculated the dispersion relations of Ag/Al$_2$O$_3$ multilayer structures with up to 8 pairs. It can be seen that the TP-TN transition occurs at 375 nm. From 1p onwards, there is no qualitative change except that, as expected, the number of branches increases progressively with the number of pairs and simultaneously the gap between the bright yellow curves fills up to tend towards forming one continuous broad band. The effective medium theory is not be the most accurate approach to predict nanophotonic enhancement, however, the literature supports that,[85] even for relatively small number of pairs, this approach is already displaying reliably HMM features as further confirmed herein.

## 3. Conclusion

In conclusion, the overall characterization of spontaneous emission properties of a range of organic emitters near HMM structures was completed by changing the number of metal-dielectric pairs as well as spacing between emitters and HMMs.

Interestingly we identified three distinct dominant regimes. These are determined based on the spontaneous emission lifetime being number of pairs independent (SPP or interference) or dependent (HD). To the best of our knowledge, this is the first report demonstrating that such a specific feature is characteristic of a HD dominant regime.

Non-monotonous variations of the spontaneous emission lifetime of organic chromophores were observed as a function of the spacer thickness which consistent with single metal film literature. Emission wavelength dependent crossovers from one dominant regime to another were discussed based on an analogy with prior single metal film investigations, and on phenomenological observations supported by wavelength and number of pairs calculation. The limit of single metal film investigation analogy was met with the dimensionless parameter $\zeta = d \cdot k_d$ which is by definition pair independent. This prevents taking advantage of this parameter to describe mulitlayer systems impact on spontaneous emission as well as to explore and describe these dominant regime crossovers. Nonetheless, it is noteworthy that the formalism developed for single metal film structure still holds even if the $p$-dependance of structure properties is not in $\zeta$ but in in the reflection coefficients. The difficulty of discriminating between SPP, HD and interference effects was discussed and potential opportunities to eventually circumvent this difficulty were presented.

We also confirmed that enhancement of emission decay rate of BBQ (and PPO) is smaller than C460 and PerDi cases due to the fact that Purcell factor is relatively high in TN HMM region compared to TP HMM region.

This work sheds new lights on semiconductor photoluminescence properties near metamaterial substrates. It will assist their use to control the spontaneous emission of molecules for fundamental and applied investigations. Based on the present fundamental study, further improvements can be realized by rationally designing and optimizing HMM structures to provide desired tunability of Purcell factor enhancement with high emission rate across broadband spectral regions.

**Experimental Section**
Chemicals: 4,4'''-bis[(2-butyloctyl)oxy] -1,1':4',1'':4'',1'''-quaterphenyl (BBQ), Coumarin 460 (C460), and 2,5-Diphenyloxazole (PPO) were purchased from Exciton. N,N-Di(1-octylheptyl)-perylene-3,4,9,10-tetra-carboxylic diimide (PerDi) was synthesized according the literature.[86] Polymethyl methacrylate (PMMA, Mw = 50 kD) and dichloromethane (DCM) were purchased from Sigma-Aldrich.

The multi-layered substrates were deposited on fused silica by Korea Advanced Nano Fab Center (KANC). The deposition was completed by ion beam evaporation to form 10 nm thick metal and oxide successive layers with a volume fraction $f_v$ = 0.5.

Steady-state absorbance and fluorescence measurements: UV-vis absorption spectra were recorded on a Hitachi U-3310 spectrophotometer. Steady-state fluorescence spectra were obtained using a Varian Cary Eclipse spectrofluorimeter.

Time-resolved photoluminescence measurements were carried out using the streak camera (Hamamatsu) with a time resolution of about 10 ps. The pulse duration was 60 fs and the repetition rate was 5 kHz. The excitation wavelengths used for our experiment was either 325 nm or 470 nm. The emission light was collected at the magic angle between the polarization of the excitation light and the polarization of the detected polarized light in order to avoid any polarization anisotropy effects. The photo-luminescence decays were fitted by a single exponential model.

Ellipsometric measurements and analyses were completed with the spectroscopic ellipsometer of rotating polarizer type from Ellipso Technology Co. Ltd.

The analytical and numerical calculation methods are described in the SI.

**Acknowledgments**
This work has been carried out in the framework of the CNRS International Associated Laboratory "Functional nanostructures: morphology, nanoelectronics and ultrafast optics" (LIA NANOFUNC). The authors are thankful to the Institut Parisien de Chimie Moléculaire, UMR 8232, Chimie des Polymères research group including Yiming Xiao, David Kreher, André-Jean Attias, Fabrice Mathevet for providing the PerDi. The authors would like to thank Jean-Charles Ribierre for early discussions, Loic Mager for contributing to the preliminary TRPL measurements used to adjust the PerDi concentration, Sang Youl Kim for commenting the ellipsometry, and Jeong Weon Wu for early discussions. KJL, YUL and PA were supported by funding of the Ministry of Science, ICT & Future Planning, Korea (201000453, 2015001948, 2014M3A6B3063706). PA would like to thank the Canon Foundation in Europe for supporting his Fellowship. The CNRS-Ewha International Research Center program has been terminated effective the 1[st] of September 2016.

**Corresponding Authors**
* e-mail: pjpandre@riken.jp, kjlee0514@ewha.ac.kr

**Author Contributions**
KJL and PA conceived and designed the experiments. KJL prepared the samples and completed both the time resolved measurements and their analysis with the feedback of PA. KJL and SJK completed and analyzed the spectroscopic ellipsometric measurements, which were discussed with PA. KJL and YUL completed the Purcell factor calculations based on invariant embedding method and FDTD, respectively. KJL and PA organized the results. KJL prepared the figures and wrote both the manuscript and SI with feedbacks from PA and YUL. All the authors could comment the manuscript.

**Additional Information**
Supplementary information is available in the on-line version of the paper:

**Notes**
The authors declare no competing financial interest.

# Supporting Information:
# Hyperbolic Dispersion Dominant Regime Identified through Spontaneous Emission Variations near Metamaterial Interfaces

Kwang Jin Lee [a,b]*, Yeon Ui Lee [a,b], Sang Jun Kim [c], Pascal André [b,c]*

[a] Department of Physics, Quantum Metamaterials Research Center, Ewha Womans University, Seoul 03760, South Korea ; [b] Department of Physics, CNRS-Ewha International Research Center, Ewha Womans University, Seoul 03760, South Korea ; [c] Ellipso Technology Co. Ltd., 358 Kwon Gwang-ro, Paldal-gu, Suwon, South Korea ; [d] Elements Chemistry Laboratory, RIKEN, Hirosawa 2-1, Wako 351-0198, Japan

**Table S1.** absorbance ($\lambda_{abs}$) and photoluminescence ($\lambda_{PL}$) of BBQ, C460 and PerDi

|  | BBQ | C460 | PerDi |
|---|---|---|---|
| $\lambda_{abs}$ max (nm) | 309 | 371 | 524 |
| $\lambda_{PL}$ max (nm) | 382 | 442 | 536 |

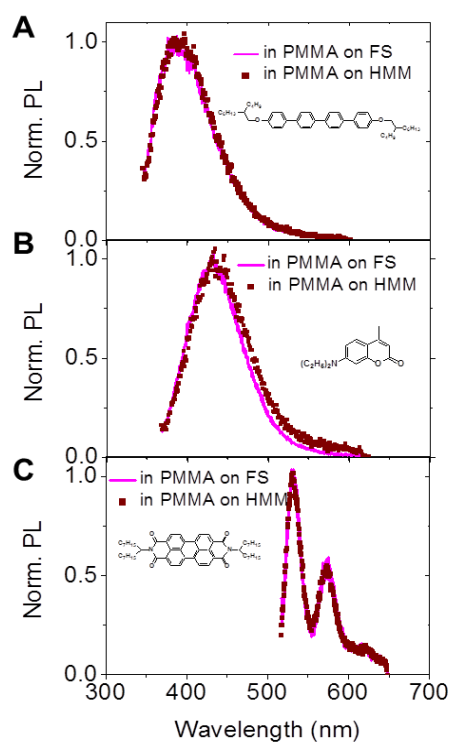

**Figure S1.** Steady state normalized fluorescence spectra: (*A*) BBQ, (*B*) C460 and (*C*) PerDi in PMMA matrix spin-coated on fused silica substrates (solid line) and on HMM substrate (filled symbol). (*A,B*) Excitation at 325 nm for BBQ and C460 and (*C*) excitation at 470 nm for PerDi.

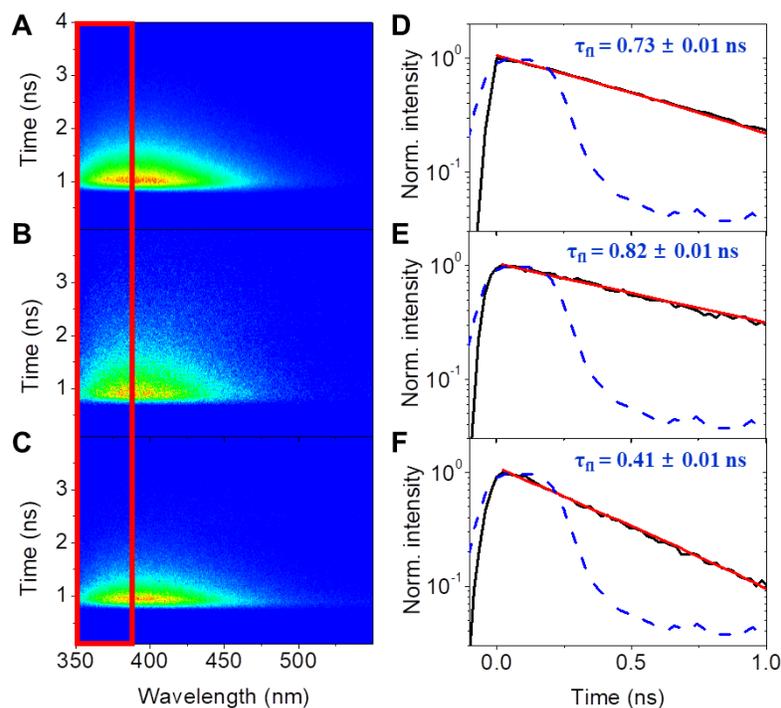

**Figure S2.** Time and spectrally resolved streak images: BBQ dispersed in (*A*) DCM, in (*B*) PMMA on fused silica and in (*C*) PMMA on a 4p Ag:$Al_2O_3$ HMM substrate with a 10 nm thick $Al_2O_3$ top cover. (*D,E,F*) Time-resolved fluorescence dynamics based on the integration of the streak image along the time axis within the boxed region ranging from 350 nm to 370 nm. Red lines are single exponential fitting lines. The dashed line is the instrument response function (IRF). The excitation wavelength is 325 nm.

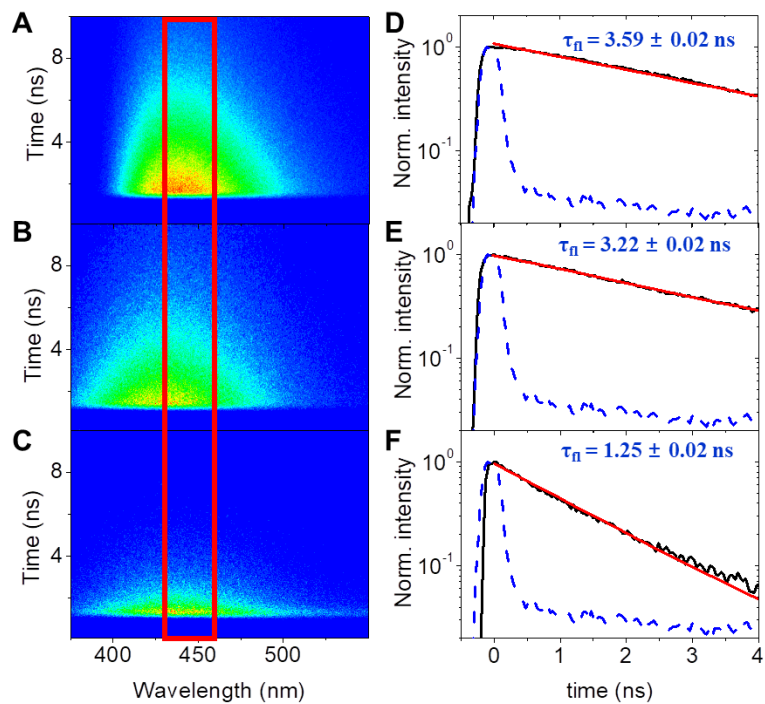

**Figure S3.** Time and spectrally resolved streak image for C460 dispersed in (*A*) DCM, in (*B*) PMMA on fused silica and in (*C*) PMMA on a 4p Ag:$Al_2O_3$ HMM substrate with a 10 nm thick $Al_2O_3$ top cover. (d,e,f) Time-resolved fluorescence dynamics based on the integration of the streak image along the time axis within the boxed regions ranging from 440 nm to 460 nm. Red lines are single exponential fitting lines. The dashed line is the instrument response function (IRF). The excitation wavelength is 325 nm.

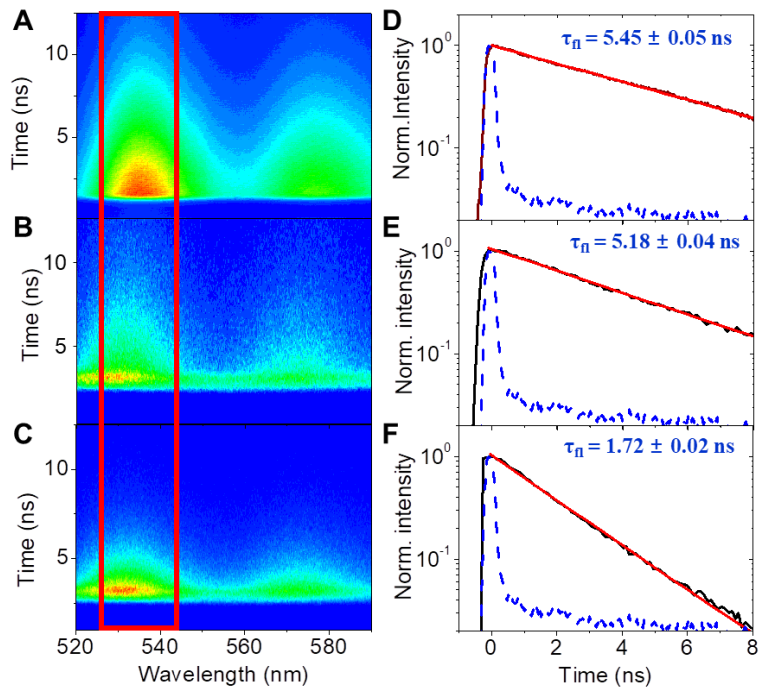

**Figure S4.** Time and spectrally resolved streak image for PerDi dispersed in (*A*) DCM, in (*B*) PMMA on fused silica and in (*C*) PMMA on a 4p Ag:Al$_2$O$_3$ HMM substrate with a 10 nm thick Al$_2$O$_3$ top cover. (*D,E,F*) Time-resolved fluorescence dynamics based on the integration of the streak image along the time axis within the boxed region ranging from 525 nm to 545 nm. Red lines are single exponential fitting lines. The dashed line is the instrument response function (IRF). The excitation wavelnght is 470 nm.

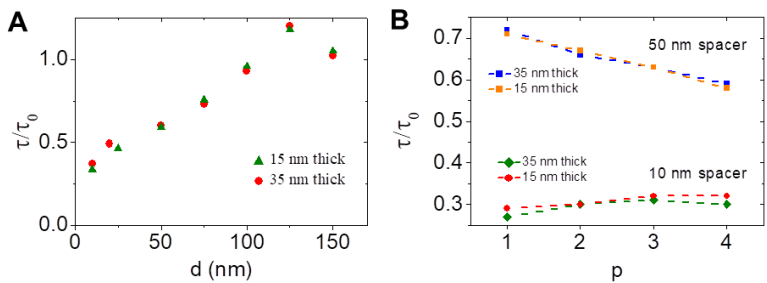

**Figure S5.** Normalized spontaneous emission lifetimes for PerDi:PMMA thin films comparing the effect of the polymer thin film thickness as a function of (*A*) distance and (*B*) number of metal-dielectric pair.

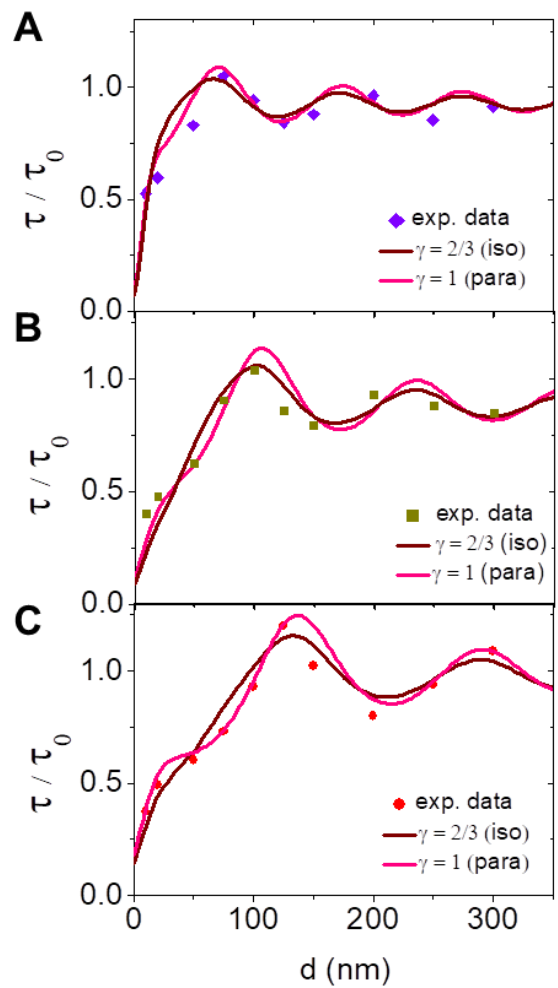

**Figure S6.** Normalized spontaneous emission lifetimes for (*A*) BBQ, (*B*) C460 and (*C*) PerDi as a function of distance up to 300 nm. Experimental data (symbol), theoretical calculations (full lines) where brown and pink lines correspond to γ=2/3 and 1, respectively.

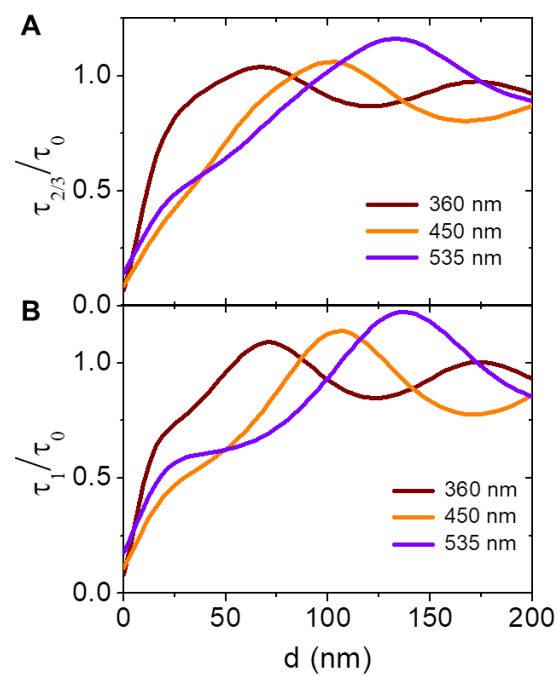

**Figure S7.** Calculation of $\tau_\gamma/\tau_0$ as a function of distance for λ = 360, 450 and 535 nm. (*A*) γ = 2/3 and (*B*) γ = 1.

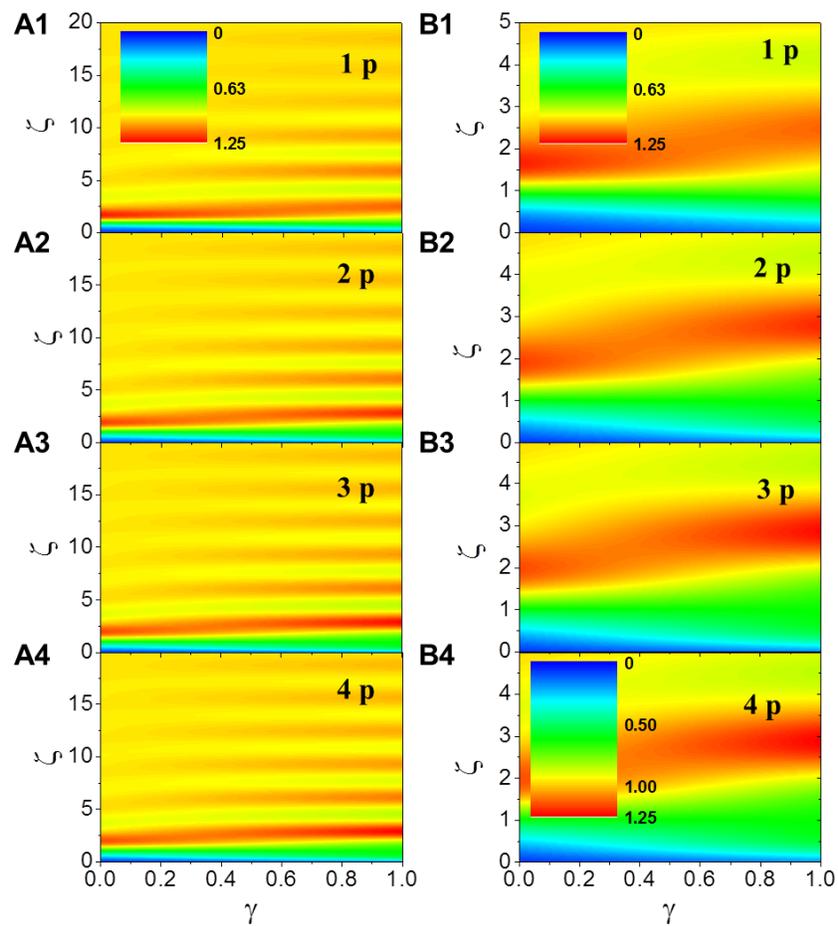

**Figure S8.** False-color plot of calculated spontaneous emission lifetime at 535 nm as a function of both dimensionless distance ($\zeta$) and orientation of the dipole ($\gamma$) for a 1 to 4 Ag/Al$_2$O$_3$ number of pairs. Up to (*A*) $\zeta$ = 20 and to (*B*) $\zeta$ = 5.

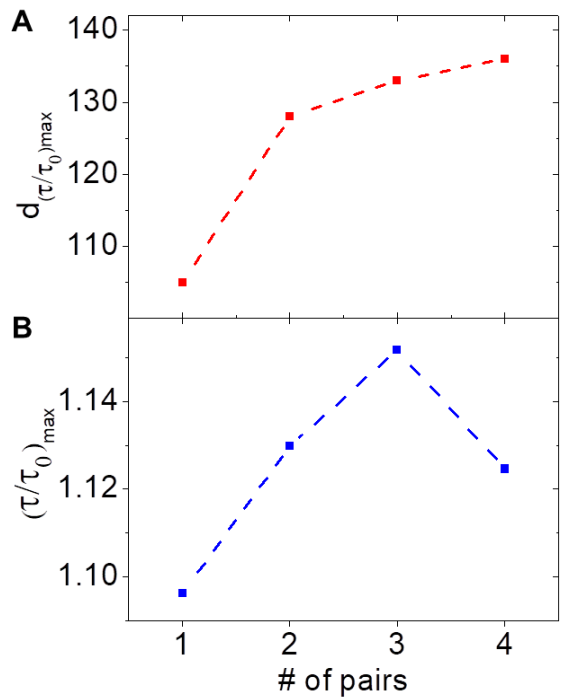

**Figure S9.** (*A*) The first maximum position ($d_{max}$) of $\tau/\tau_0$ and (*B*) the maximum value of $\tau/\tau_0$ as a function the number of Ag/Al$_2$O$_3$ pairs at $\lambda = 535$ nm, respectively.

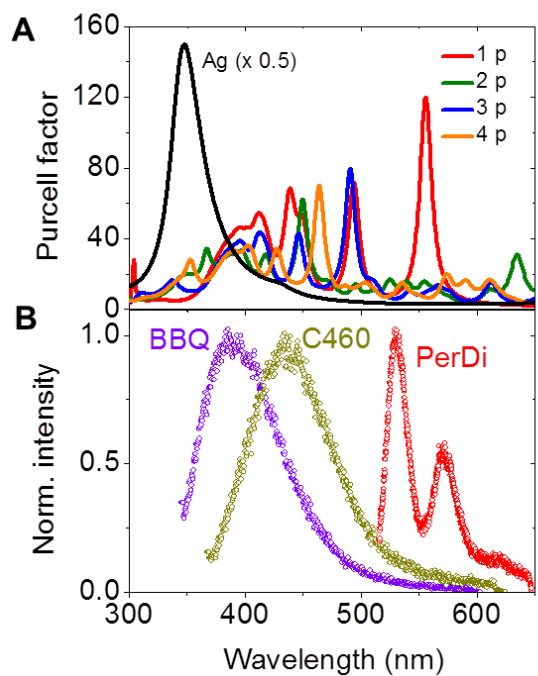

**Figure S10.** (*A*) Purcell factor calculated for single Ag film (70 nm thick without Al$_2$O$_3$) and for 1 to 4 Ag/Al$_2$O$_3$ pairs (10 nm thick) substrates (*B*) Steady state normalized fluorescence for BBQ, C460 and PerDi in PMMA matrix spin-coated on fused silica substrates.

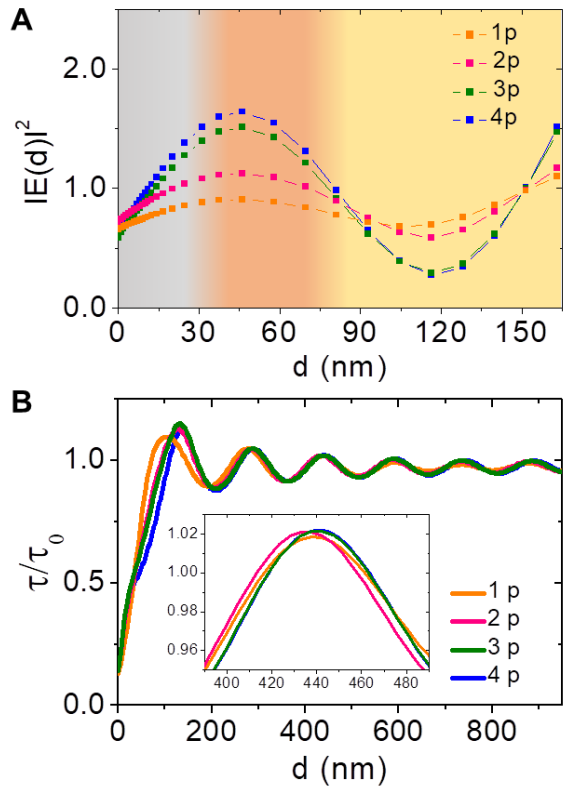

**Figure S11.** (*A*) FDTD calculation of the optical field as a function the spacer thickness and (*B*) calculated $\tau/\tau_0$ oscillations as a function the spacer thickness and the number of Ag/Al$_2$O$_3$ pairs, at $\lambda$ = 535 nm.

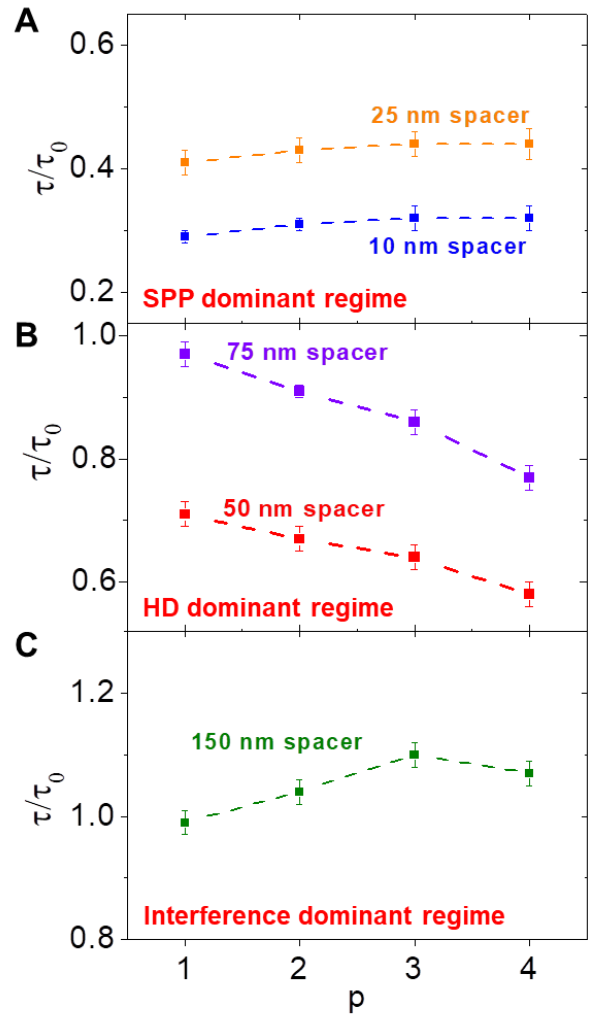

**Figure S12.** Experimental data of the emission lifetime ratio of transient dipoles at 535 nm as a function of the number of pairs of the HMM structure : (*A*) Surface Plasmon Polariton, (*B*) Hyperbolique Dispection, and (*C*) interference regimes.

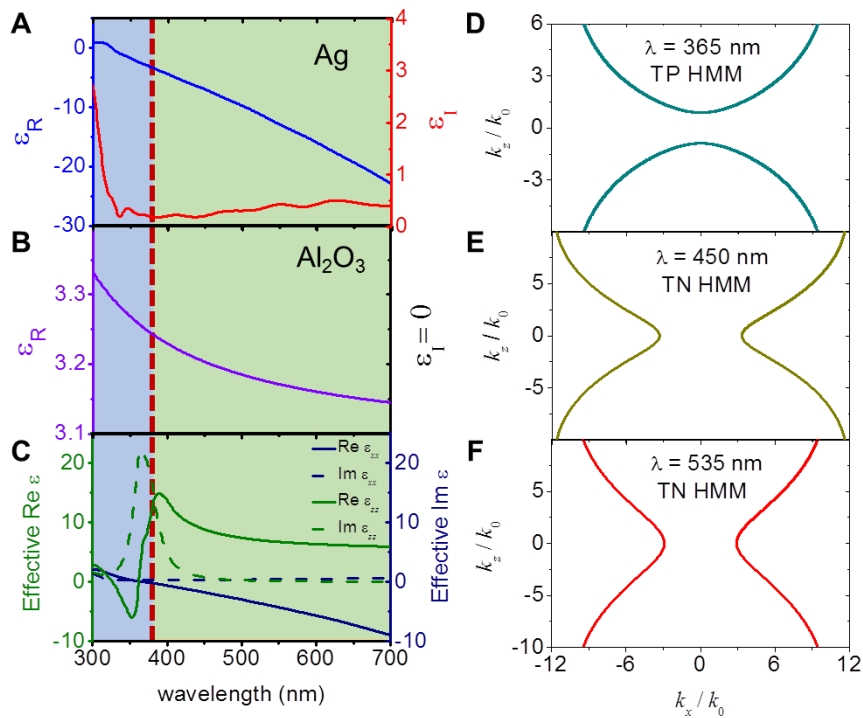

**Figure S13.** Dielectric constants of (*A*) Ag and (*B*) Al$_2$O$_3$, and (*C*) effective dielectric constants for both real and imaginary part based on Eq. (S6) and (S7). Dispersion relation calculation based on Eq.(S8) and for infinite Ag/Al$_2$O$_3$ multilayers at (*D*) 365 nm, (*E*) 450 nm and (*F*) 535 nm wavelengths, respectively.

Dielectric constants of Ag and Al$_2$O$_3$ shown in Figure S13*A-B* are obtained by spectroscopic ellisometry measurements. Dispersion relations of infinite multilayer HMM for three wavelengths are calculated and plotted in Figure S13*D-E-F* based on Bloch theorem.

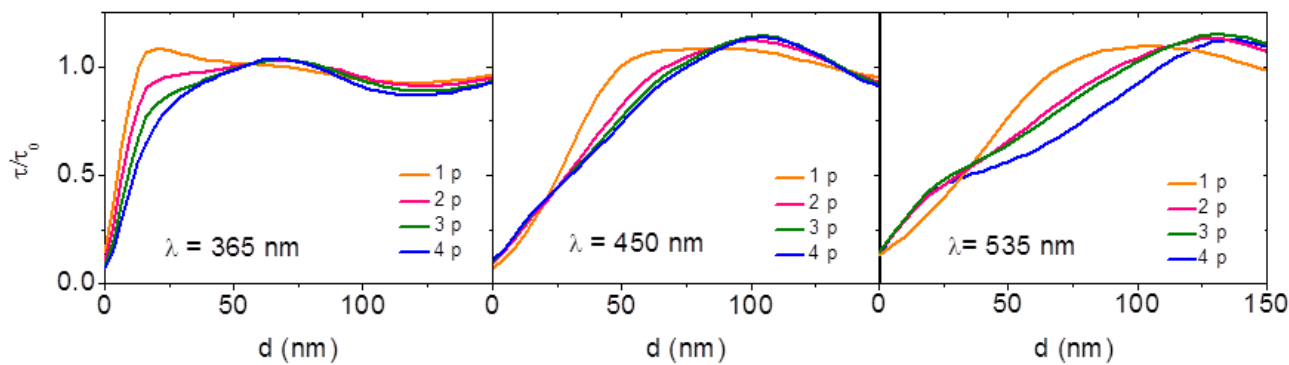

**Figure S14.** Calculations of $\tau/\tau_0$ as a function of distance from 1 ~ 4 Ag/Al2O3 pairs at $\lambda$ =365, 450 and 535 nm.

We also definitely expect that $\tau/\tau_0$ values of BBQ and C460 at $d$ = 10 nm for 1~4 pairs are not much different since (1) Purcell factors at $\lambda$ = 365 and 450 nm for 1~4 pairs show much smaller deviations and (2) emission band width of C460 or BBQ is even broader than that of PerDi as shown in Figure S8*A*. In Figure S9, the same calculation of $\tau/\tau_0$ at 450 nm for 1 ~ 4 pairs is displayed and almost similar values of $\tau/\tau_0$ for small distance compared to 535 nm. In other words, high $k_{nr}$ results in shorter $\tau$ in this close range since the dipole inside chromophore is mostly influenced by top metal surface regardless of the number of metal-dielectric pairs.

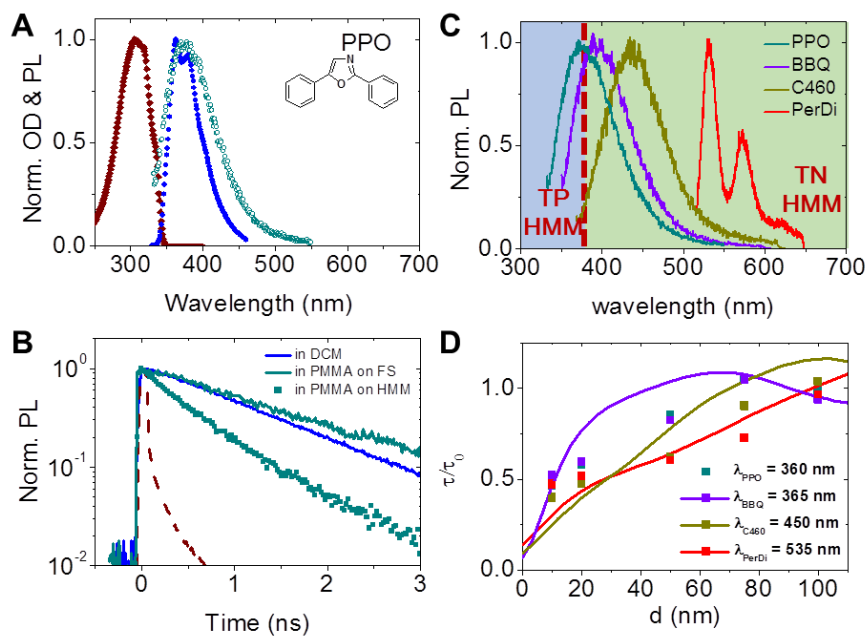

**Figure S15.** (*A*) Steady state normalized absorbance (Abs., ♦) and photoluminescence (PL, ●) for PPO in DCM solutions (filled symbol), in PMMA matrix spin-coated on a fused silica (FS) substrate (empty symbol). (*B*) Time-resolved photoluminescence of PPO in DCM solution, and in PMMA films spincoated on FS and HMM substrates. The dashed line is the instrument response function. (*C*) Fluorescence spectra of four chromophores including PPO dispersed in PMMA. The dashed line is the boundary wavelength between transverse positive (TP) and transverse negative (TN) $Ag:Al_2O_3$ HMMs. (*D*) Calculated (solid lines) of normalized spontaneous emission lifetimes for λ = 360, 365, 450 and 535 nm and experimental data (symbols) for four chromophores.

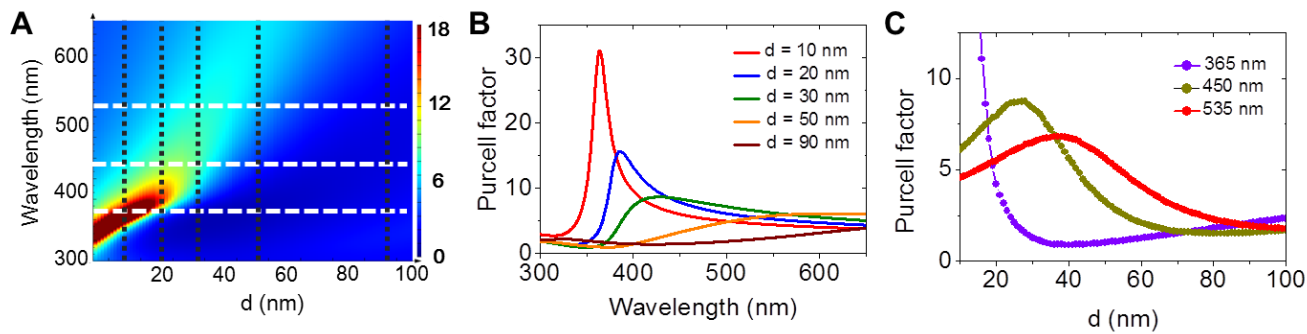

**Figure S16.** (*A*) False-color plot of Purcell factor as a function of wavelength and distance from a 70 nm thick Ag film. (*B*) Cross-sections of the Purcell factor as a function of wavelength taken at d = 10, 20, 30, 50 and 90 nm materialised by the vertical black dotted lines in Figure S16. (*C*) Purcell factor as a function of distance for three wavelengths corresponding to horizontal white dashed lines in Figure 16*A*.

Figure S16, corresponds to the Purcell factor calculations obtained for a single 70 nm thick Ag film substrate. Comparing Figure 5D and Figure S16A shows that the Purcell factor in the presence of HMM presents a broadband frequency tunability as a function of distance *d* and this demonstrates that the broadband Purcell peak results from the higher number of hybridized plasmonic modes made available to the emitter. This is shown in Figure S16B and C.

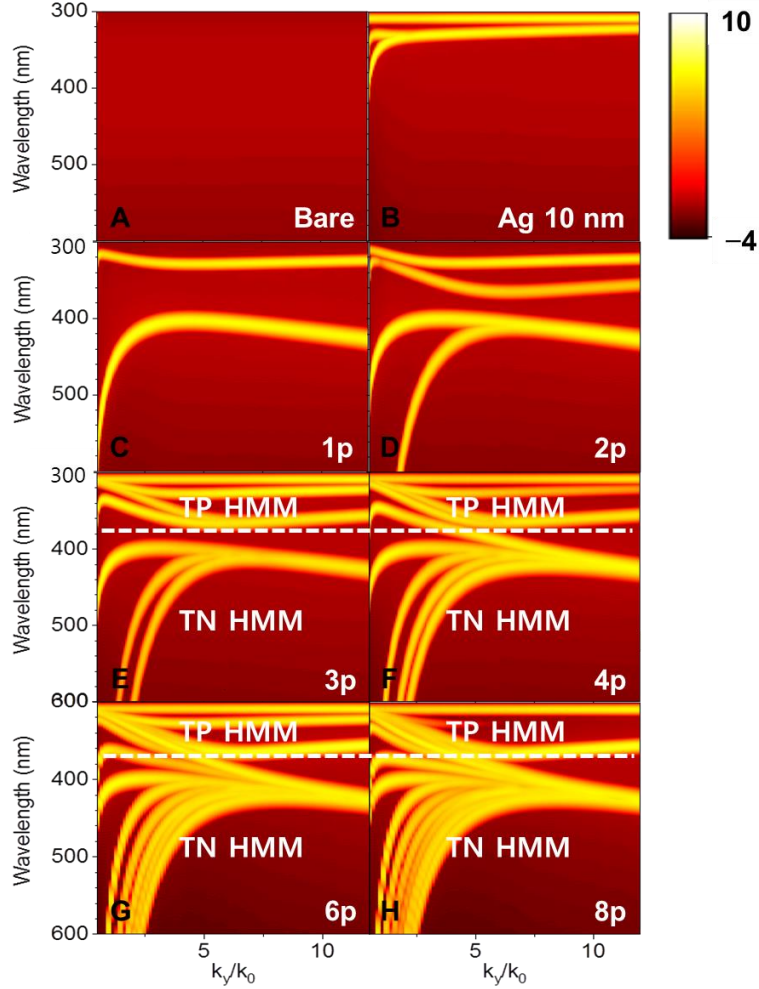

**Figure S17.** Dissipated power spectra. (*A*) air, (*B*) single Ag film, (*C*) 1 Ag/Al$_2$O$_3$ pair, (*D*) 2 Ag/Al$_2$O$_3$ pairs, (*E*) 3 Ag/Al$_2$O$_3$ pairs, (*F*) 4 Ag/Al$_2$O$_3$ pairs, (*G*) 6 Ag/Al$_2$O$_3$ pairs, (*H*) 8 Ag/Al$_2$O$_3$ pairs.

## Spectroscopic Ellipsometry

The rotating polarizer type spectroscopic ellipsometer (Ellipso Technology Co. Ltd.) is used in the present measurements. These are performed in the spectral range of 1.2–5.2 eV (240-1000 nm) for three angles of incidence from 60° to 70° with a 5° step. This approach improves the accuracy of modelling analysis, allowing determination of film thicknesses and refractive index values. The measured ellipsometric constants $\Psi$ and $\Delta$ are defined from the ratio of the reflection coefficients $r_p$ and $r_s$ for the p- and s- polarizations, respectively, (i.e., electric field parallel and perpendicular to the plane of incidence) according to the following equation.

$$\tan\Psi e^{i\Delta} = \frac{r_p}{r_s} \tag{S1}$$

From the analysis of the measured SE data, the dielectric function of the material of interest is derived. The complex dielectric function of the polymer films like BBQ, C460, and PerDi is expressed by the Tauc-Lorentz dispersion formula,[1] allowing multiple oscillators.

$$\varepsilon_{TL}(E) = \varepsilon_r(E) + i\,\varepsilon_i(E) \tag{S2}$$

$$\varepsilon_i(E) = \begin{cases} \frac{1}{E} \frac{AE_oC(E-E_g)^2}{(E^2-E_0^2)^2+C^2E^2} & for\ E > E_g \\ 0 & for\ E \leq E_g \end{cases} \qquad (S3)$$

$$\varepsilon_r(E) = \varepsilon_r(\infty) + \frac{2}{\pi} \cdot P \int_{E_g}^{\infty} \frac{\xi \cdot \varepsilon_i(\xi)}{\xi^2 - E^2} d\xi \qquad (S4)$$

The coefficients of above dispersion equation are determined by fitting the calculated ellipsometric constants to the measured ones, that is, by minimizing the error function F, presented in equation S5. The Levenberg-Marquardt algorithm is used for fast convergence.

$$F = \frac{1}{2N-M} \sum_{i=1}^{N} \left[ \left(\alpha_i^{cal} - \alpha_i^{exp}\right)^2 + \left(\beta_i^{cal} - \beta_i^{exp}\right)^2 \right] \qquad (S5)$$

Here $N$ is the number of $(\alpha, \beta)$ pairs, and $M$ is the number of fitting parameters. The superscripts *cal* and *exp* indicate calculated and measured, respectively.

## Theoretical methods

1. Effective dielectric functions

Using the effective medium theory, the effective dielectric functions of transverse components $\varepsilon_\parallel$ and that of longitudinal component $\varepsilon_\perp$ are expressed as follow.

$$\varepsilon_\parallel = p\varepsilon_m + (1-p)\varepsilon_d \qquad (S6)$$

$$\varepsilon_\perp = \frac{\varepsilon_m \varepsilon_d}{p\varepsilon_d + (1-p)\varepsilon_m} \qquad (S7)$$

Here $\varepsilon_m$ and $\varepsilon_d$ are the dielectric functions of metal (silver, Ag) and that of the dielectric (alumina, $Al_2O_3$), respectively. $p$ is the filling ratio of the metal. Our case, $p$ is equal to 0.5.

2. Dispersion relation calculation

Since the multilayer is periodic in z-axis, it supports propagating Bloch waves with the wavevector $k^2 = k_B^2 + k_x^2$, the relation between $k_B$ and $k_x$ is given by the Bloch theorem,

$$\cos[k_B(t_m + t_d)] = \cos(u_m t_m) \cos(u_d t_d) - [(\beta + \beta^{-1})/2] \sin(u_m t_m) \sin(u_d t_d) \qquad (S8)$$

where $u_j = [\varepsilon_j \omega^2/c^2 - k_x^2]^{1/2}$ ($j=m,d$) is the z-component of the wavevector in metal or dielectric and $\beta = (\varepsilon_m u_d)/(\varepsilon_d u_m)$. $t_m$ and $t_d$ are the thickness of metal and dielectric, respectively. In Figure S5, $k_B$ corresponds to $k_z$.

3. Invariant imbedding method

Invariant imbedding method is used for electromagnetic wave propagation in inhomogeneous medium.

We are interested in the propagation of a monochromatic electromagnetic wave of angular frequency $\omega$ and vacuum wave number $k_0 = \omega/c$, where $c$ is the speed of light in vacuum. The wave is assumed to be incident from a homogeneous region on a layered or stratified medium, where the dielectric permeability $\varepsilon$ varies only in one-direction in space. We take this direction as the z-axis and assume the inhomogeneous, but isotropic, medium lies in $0 \leq z \leq L$. Without loss of generality, we assume that the wave propagates in the xz-plane. Since the medium is uniform in the x-direction, the dependence on x can be taken as being through a factor $e^{ik_x x}$.
We assume that the wave is incident from the region where $z > L$ and is transmitted to the region where $z < 0$. The dielectric permittivity is assumed to be given by

$$\varepsilon(z) = \begin{cases} \varepsilon_1 & \text{for } z > L \\ \varepsilon_L(z) & \text{for } 0 \leq z \leq L \\ \varepsilon_2 & \text{for } z < 0 \end{cases} \quad (S9)$$

where $\varepsilon_1$ and $\varepsilon_2$ correspond to real dielectric constants of incident ($Al_2O_3$ layer) and transmitted (glass, FS) region and $\varepsilon_L(z)$ corresponds to arbitrary real dielectric function of z in HMMs.

We will consider electric field given by

$$E(x,z) = E(z) e^{ik_x x} = e^{k_1[-w(L-z)+iux]} \quad (S10)$$

where $u = k_x/k_1$, $w = -i(1-u^2)^{1/2}$. $k_1$ is the wavevector in incident region given by $nk_0 = 2\pi n/\lambda$ where $n$ is the refractive index of incident region. Based on the above relation, we can obtain the electric field in the incident region

$$E(x,z) = \left[ e^{-k_1 w(L-z)} + r(L) e^{-k_1 w(z-L)} \right] e^{ik_1 ux} \quad (S11)$$

where $r$ is the reflection coefficient.

Using the invariant imbedding method,[2, 3] we can obtain the first order differential equations for reflection coefficients of s- and p- polarized waves denoted by $r_s$ and $r_p$, respectively,

$$\frac{1}{k_1} \frac{dr_s(l)}{dl} = -2wr(l) - \frac{w}{2(1-u^2)}[\varepsilon(l) - 1][1 + r(l)]^2 \quad (S12)$$

$$\frac{1}{k_1} \frac{dr_p(l)}{dl} = -2w\varepsilon(l)r(l) - \frac{w}{2}\left[ \varepsilon(l) + \frac{u^2}{\varepsilon(l)(1-u^2)} + \frac{1}{(1-u^2)} \right][1 + r(l)]^2 \quad (S13)$$

where $l$ is the inhomogeneous medium size parameter varying from 0 to $L$. i.e. from the fused silica/HMM interface to the $Al_2O_3$ / PMMA thin film interface.

4. Purcell factor

Finite-difference time-domain (FDTD, Lumerical)[4, 5] simulations were performed to obtain the Purcell factor. The Purcell factor is the emission rate enhancement of a spontaneous emitter inside/near a cavity or plasmonic structure, which can be obtained directly from the FDTD simulations. By placing a dipole source on top of HMM or Ag film at $z = d$, the dividing the power emitted from a dipole in the presence of HMM by the power emitted from the dipole in the absence of HMM is calculated upon a 300-700nm. The geometries of the HMM structure in the simulations were designed to match with the Figure 3A. A minimum mesh step size of 0.25 nm was defined, and the perfectly matched layers boundary conditions are adopted. For the permittivity of the glass substrate we employed the values obtained from ellipsometer measurement, and the permittivity of Ag is adapted from the CRC handbook. The calculation of Purcell factor includes both radiative and non-radiative decay rates for the dipole modifying from plasmonic structures adjacent to the dipole.